\documentclass[12pt,preprint]{aastex}
\usepackage{epsfig}
\usepackage{lscape}

\makeatletter

\newcommand{\Rmnum}[1]{\expandafter\@slowromancap\romannumeral #1@}
\makeatother

\def\Msun{\hbox{\it M$_\odot$}}

\newcommand{\program}[1]{{\tt {#1}}}
\def\program{\texttt}
\def\ca{\citeauthor}
\def\cy{\citeyear}

\begin{document}
\title{Discovery of Twin Wolf-Rayet Stars Powering Double Ring Nebulae}

\shorttitle{}
\author{Jon C. Mauerhan\altaffilmark{1}, Stefanie Wachter \altaffilmark{1}, Patrick W. Morris\altaffilmark{2}, Schuyler D. Van Dyk \altaffilmark{1}, and D. W. Hoard\altaffilmark{1}}
\altaffiltext{1}{Spitzer Science Center, California Institute of Technology, Mail Code 220-6, 1200 East California Blvd., Pasadena, CA 91125, USA; mauerhan@ipac.caltech.edu} 
\altaffiltext{2}{NASA Herschel Science Center, California Institute of Technology, Pasadena, CA 91125, USA} 

\begin{abstract}
We have spectroscopically discovered a pair of twin, nitrogen-type, hydrogen-rich, Wolf-Rayet stars (WN8--9h) that are both surrounded by circular, mid-infrared-bright nebulae detected with the \textit{Spitzer Space Telescope} and MIPS instrument.  The emission is probably dominated by a thermal continuum from cool dust, but also may contain contributions from atomic line emission. There is no counterpart at shorter \textit{Spitzer}/IRAC wavelengths, indicating a lack of emission from warm dust. The two nebulae are probably wind-swept stellar ejecta released by the central stars during a prior evolutionary phase. The nebulae partially overlap on the sky and we speculate on the possibility that they are in the early stage of a collision. Two other evolved massive stars have also been identified within the area subtended by the nebulae, including a carbon-type Wolf-Rayet star (WC8) and an O7--8 III--I star, the latter of which appears to be embedded in one of the larger WN8--9h nebulae. The derived distances to these stars imply that they are coeval members of an association lying $4.9\pm1.2$ kpc from Earth, near the intersection of the Galaxy's Long Bar and the Scutum-Centaurus spiral arm. This new association represents an unprecedented display of complex interactions between multiple stellar winds, outflows, and the radiation fields of evolved massive stars.
\end{abstract}

\section{Introduction}
Massive stars are key in the mechanical and chemical evolution of galaxies. As sources of ionization, kinetic energy, and heavy metals, massive stars have a profound influence on the interstellar medium (ISM), spurring and regulating star formation, and seeding the ISM with the necessary ingredients for the formation of planets and life. Yet despite their prominent role in cosmic evolution, our understanding of massive stars is sparse, particularly regarding the post main-sequence stages where they enter a  phase of heavy mass loss, shed their hydrogen envelopes, and become Wolf-Rayet (WR) stars. Although we know that mass loss via steady winds may be punctuated by large discrete ejections during the Luminous Blue Variable (LBV) stage, we do not know what triggers such eruptions or what supplies their energy. This is a major limitation of stellar evolutionary models, since it is thought that the majority of a massive starÕs hydrogen envelope is shed through such eruptions (\ca{langer94}~\cy{langer94}; \ca{cro95}~\cy{cro95}, \cy{cro07}, and references therein). For nearly all massive stars, we will never witness these eruptive events. However, the ejections create lasting circumstellar nebulae, whose morphologies, kinematics, and abundances provide the best physical probe of the mass-loss history.

Recently, the \textit{Spitzer Space Telescope} and Mulitband Imaging Photometer for Spitzer (MIPS; \ca{rieke04}~\cy{rieke04}) were used to conduct the MIPSGAL Legacy survey (\ca{car09}~\cy{car09}), which imaged the inner Galactic plane at $\lambda$24 {\micron} and $\lambda$70 {\micron}. \ca{miz10}~(\cy{miz10}) catalogued 416 disk-like and ring-like nebular structures detected in this survey. \ca{wac10}~(\cy{wac10}) compiled a more restricted sample of circular and elliptical nebulae that surround central point sources, many of which are bright at near-infrared and optical wavelengths. Our ongoing spectroscopic survey to classify the central objects of these nebulae has resulted in the identification of dozens of rare, evolved massive stars, including new candidate LBVs, WRs, and other OB and emission-line stars. This sample represents a significant observational advance in massive-star astronomy that will yield valuable insights into the nature of discrete mass ejections from massive stars, as well as the evolutionary connection between the various spectroscopic morphologies that massive stars exhibit, from their departure off the main-sequence to their eventual core collapse. 

Two particularly interesting nebulae within the MIPSGAL sample are MGE028.4812$+$00.3368 and MGE028.4451$+$00.3094 (nomenclature from \ca{miz10}~\cy{miz10}), shown in Figure 1. These sources correspond to Shells \#47 and \#48 from \ca{wac10}~\cy{wac10}, and MN85 and MN86 from \ca{gvar10}~(\cy{gvar10}), respectively. Both nebulae are circular and are centered on reddened point sources that happen to be candidate WR stars, according to the infrared color criteria of \ca{had07}~(\cy{had07}) and \ca{mau09}~(\cy{mau09}). We call these stars WMD47$^*$ and WMD48$^*$, after their designations in \ca{wac10}~(\cy{wac10}); they correspond to stars 2MASSJ 18420630$-$0348224 and 2MASSJ 18420827$-$0351029, respectively. Furthermore, the area subtended by MGE028.4812$+$00.3368 and MGE028.4451$+$00.3094 contains two other stars of interest, in addition to WMD47$^*$ and WMD48$^*$. The star 2MASSJ 18420846$-$0349352 is also a candidate WR, according to the infrared color criterion referenced above, and 2MASSJ 18420957$-$0350313 lies near the center of curvature of a bright ridge of $\lambda$24 {\micron} emission, approximately halfway from the center to the rim of MGE028.4451$+$00.3094. The ridge appears to be the site of intersection of stellar winds. Based on the collective characteristics of the nebulae, the candidate massive stars in this region were given priority in our spectroscopic survey. In this Letter, we present the results of our spectroscopic investigation of these massive stars and speculate on their connection to the dual nebulae. 

\section{Infrared Spectroscopy}
Infrared spectra of 2MASSJ 18420630$-$0348224 (WMD47$^*$), 2MASSJ 18420827$-$0351029 (WMD48$^*$), 2MASSJ 18420957$-$0350313 and 2MASSJ 18420846$-$0349352  were obtained with the Palomar Hale 5 m telescope and the Triplespec (TSPEC; \ca{hert08}~\cy{hert08}) near-infrared spectrograph on 2010 June 20 UT. TSPEC provides simultaneous coverage of the $J$, $H$, and $K$ bands ($\lambda=1.0$--2.4 {\micron}) and produces a moderate resolution spectrum ($R\approx2500$--2700) through a $1\arcsec \times 30\arcsec$ slit. Spectra exposures were obtained in an ABBA telescope nodding sequence for sky subtraction and bad-pixel suppression. Wavelength calibration was performed using the OH emission lines in the sky spectra and flat-field calibration was achieved using spectra of continuum lamps in the dome. Spectra of A0V stars were obtained to derive a telluric absorption spectrum, which was produced and applied to the science data using the \program{IDL}-based program \textit{xtellcor} (\ca{vac03}~\cy{vac03}).

\subsection{WMD47$^*$ and WMD48$^*$: The Twin WNh Stars}
The remarkably similar near-infrared spectra of  WMD47$^*$ and WMD48$^*$ are presented in Figure 2.  The $K$-band is dominated by strong He {\sc i} emission  at $\lambda$2.058 {\micron} and $\lambda$2.112 {\micron}, and H~{\sc i} at $\lambda$2.166 {\micron} (Br$\gamma$). He {\sc i} is also detected as a weak emission feature at $\lambda$2.185 {\micron}; this feature likely contains contributions from two He~{\sc i} 7p--4d transitions at $\lambda$2.1821 {\micron} (triplet)  and $\lambda$2.1846 {\micron} (singlet). He~{\sc ii} emission at $\lambda$2.189 {\micron} appears to be slightly blended with, yet clearly separable from, the He~{\sc i} feature. Weak emission from N~{\sc iii} is also present at $\lambda$2.247 {\micron}. In the $H$ band, these stars exhibit strong He~{\sc i} emission at $\lambda$1.700 {\micron} and weak He~{\sc ii} emission at $\lambda$1.692 {\micron}, in addition to the prominent Brackett series of H~{\sc i}. Finally, the $J$-band spectra are dominated by a very strong P-Cygni feature of He~{\sc i} at $\lambda$1.083 {\micron} and Paschen-$\beta$ at $\lambda$1.28 {\micron}.  All of the above spectral features are characteristic of late nitrogen-type WR stars (\ca{pmor96}~\cy{pmor96}; \ca{fig97}~\cy{fig97}). A specific WN subtype diagnostic for these types is provided by the equivalent-width  (EW) ratio of He~{\sc ii} at $\lambda$2.189 {\micron} relative to Br$\gamma$ and He~{\sc i} at $\lambda$2.112 {\micron} (\ca{fig97}~\cy{fig97}; \ca{cro06}~\cy{cro06}). We measured EW for He~{\sc ii} by using the \textit{deblend} feature in \program{IRAF}, which fit overlapping gaussian profiles for the weak He~{\sc i} and He~{\sc ii} lines. WMD47$^*$ and WMD48$^*$ have EW(He~{\sc ii} 2.189)/EW(Br$\gamma$)$\approx$0.03 and EW(He~{\sc ii 2.189})/EW(He~{\sc i} 2.112)$\approx$0.07--0.08, which is consistent with the WN8--9 subtype range. The prominent Brackett emission series of H {\sc i} in the $H$-band indicates that the envelopes of WMD47$^*$ and WMD48$^*$ are hydrogen rich. As such, they may be classified as WN8--9h stars. 

\subsection{2MASSJ 18420846$-$0349352 and 2MASSJ 18420957$-$0350313}
The spectrum of 2MASSJ 18420846$-$0349352 is presented in Figure 3. In the $K$-band, this star exhibits very strong and broad emission lines of C~{\sc iii} and C~{\sc iv} near $\lambda$2.07--2.08 {\micron}, C~{\sc iii} at $\lambda$2.112--2.115 {\micron}, He~{\sc i} and He~{\sc ii} near $\lambda$2.058 \micron, and He~{\sc ii} near $\lambda$2.189 {\micron}. These are spectral characteristics of late carbon-type (WC) WRs. A subtype diagnostic is provided by the EW ratio of C~{\sc iv} at $\lambda$2.08 {\micron} relative to C~{\sc iii} at $\lambda$2.112 {\micron} (\ca{fig97}~\cy{fig97}; \ca{cro06}~\cy{cro06}), which has a value of $\approx$2 for this star. This is consistent with WC8 subtype. 

The spectrum of 2MASSJ 18420957$-$0350313 is also shown in Figure 3. It exhibits H~{\sc i} (Br$\gamma$) and He~{\sc i} in absorption, and weak emission features of N~{\sc iii} at $\lambda$2.115 {\micron} and $\lambda$2.247 {\micron}, the former of which lies on the red side of the He~{\sc i} absorption line at $\lambda$2.112 {\micron}, not to be mistaken for a P-Cygni profile. The Br$\gamma$ absorption line appears to be partially filled with an emission component, which is probably produced by the stellar wind. There is also a He~{\sc ii} absorption feature at $\lambda$2.189 {\micron}. The spectrum bears strong resemblance to the O7--8 V--I stars in the atlases by \ca{han96}~(\cy{han96, han05}); so, we classify 2MASSJ 18420957$-$0350313 as an O7--8 star. Unfortunately, the luminosity class of this star is not effectively constrained by the spectral morphology. We will address the topic of luminosity in Section 3, using photometry.

\section{Extinction and Distance}
The $JHK_s$ photometry for each star is presented in Table 1. We computed the interstellar extinction for each source by combining the measured photometry with intrinsic $J-K_s$ and $H-K_s$ colors and $M_{K}$ values that were adopted from the literature (\ca{cro06}~\cy{cro06}; \ca{mar06}~\cy{mar06}).  After the extinction-induced color excesses ($E_{J-K_s}$ and $H_{H-K_s}$) were derived, we used the extinction relation of \ca{ind05}~(\cy{ind05}) to calculate two values of $A_{K_s}$, given by \[  A_{K_{s}} = 1.82^{+0.30}_{-0.23} E_{H-K_{s}} \] and \[  A_{K_{s}} = 0.67_{-0.06}^{+0.07} E_{J-K_{s}} \] where $E(J-K_{s})=(J-K_{s})_{\textrm{\scriptsize{obs}}}-(J-K_{s})_{0}$, and similarly for the $H-K_s$ colors. The two results were averaged to obtain a final $\overline{A_{K_s}}$ estimate.  Using the adopted values of $M_K$, we then derived the distance to each source. The results are presented in Table 1, along with all of the adopted values that we used from the literature.  

The distances to the three WRs imply that they are all members of the same physical association and are, hence, coeval. We used their combined average distance of the $4.9\pm1.2$ kpc to derive a value of $M_K=-5.0\pm0.6$ mag for the O7--8 star 2MASSJ 18420957$-$0350313. According to \ca{mar06}~(\cy{mar06}), O7--8 III stars should have $M_K$ between $-$4.76 and $-$4.62 mag, while O7--8 I stars should have $M_K\approx-5.5$ mag,  The value of $M_K=-5.0\pm0.6$ mag that we obtain for 2MASSJ 18420957$-$0350313 is intermediate; however, given the estimated uncertainty, our data cannot constrain the luminosity class any tighter than the range of III--I. 

The angular radii of MGE028.4812$+$00.3368 and MGE028.4451$+$00.3094 and the derived distance implies physical radii of $\approx$2.5 and $\approx$2.1 pc, respectively. The location of the nebulae at $(l,  b)=(28.46^{\circ}, 0.32^{\circ})$ implies that this system lies near the intersection of the Galaxy's Long Bar and the Scutum-Centaurus spiral arm (\ca{church09}~\cy{church09}), a.k.a.~the Scutum-Crux arm. This general area of the Galaxy is known to be the site of vigorous massive star formation, as indicated by the recent discovery of four massive clusters of red supergiants between $l=24^\circ$--$29^\circ$ (\ca{fig06}~\cy{fig06}; \ca{dav07}~\cy{dav07}; \ca{al09}~\cy{al09}; \ca{neg10}~\cy{neg10}). 

\section{Discussion}
The similar spectral morphologies of WMD47$^*$ and WMD48$^*$ and presence of circumstellar nebulae around them suggests that these stars are at practically identical evolutionary stages. WN8--9h stars surrounded by such nebulae are believed to have recently been through an abrupt stage of heavy mass loss, i.e., an LBV or red supergiant (RSG) phase (\ca{cro95}~\cy{cro95}). The M1-67 nebula surrounding the WN8h star WR124 is a particularly well known example of this phenomenon (\ca{est91}~\cy{est91}; \ca{gros98}~\cy{gros98}).  As the central stars of such systems enter the WR stage, they develop supersonic winds having $V_{\textrm{\tiny{WR}}}\gtrsim1000$ km s$^{-1}$, which sweep up the slower moving mass lost in the prior RSG or LBV phases ($V_{\textrm{\tiny{RSG}}}\sim10$ km s$^{-1}$; $V_{\textrm{\tiny{LBV}}}\sim100$ km s$^{-1}$).  The expanding shell that forms will have a velocity given by $V_{\textrm{\tiny{shell}}}\sim(\dot{M}_{\textrm{\tiny{WR}}} V_{\textrm{\tiny{WR}}}^2 V_{\textrm{\tiny{ej}}}/3\dot{M}_{\textrm{\tiny{ej}}})^{1/3}$ (\ca{chev83}~\cy{chev83}), where the $\dot{M}$ and $V$ factors each refer to the mass-loss rates and expansion velocities of the WR wind and of the ejecta from the RSG or LBV precursor, respectively. Assuming that the WN8--9h wind has $v\sim1000$ km s$^{-1}$ and $\dot{M}\sim10^{-5}$ {\Msun} yr$^{-1}$ (\ca{mar08}~\cy{mar08}), the ejecta from either a RSG ($\dot{M} \sim 3\times10^{-5}$ {\Msun} yr$^{-1}$) or LBV ($\dot{M} \sim10^{-3}$ {\Msun} yr$^{-1}$) progenitor will achieve a shell velocity of $\sim$100 km s$^{-1}$. Using the observed radii of $\approx$2.1 pc and $\approx$2.5 pc  for MGE028.4812$+$00.3368 and MGE028.4451$+$00.3094, respectively, implies shell ages of $\approx2$--$2.5\times10^4$ yr. This is comparable to the  dynamical time scales of M1-67 (\ca{est91}~\cy{est91}) and the nebula surrounding the WN8--9h star WR138a (\ca{gvar09}~\cy{gvar09}), the latter of which has a circular, shell-like morphology that is very similar to both MGE028.4812$+$00.3368 and MGE028.4451$+$00.3094.

\subsection{Speculations on the Origin of the Nebular Emission and Morphology}
MGE028.4812$+$00.3368 and MGE028.4451$+$00.3094 are primarily detected in the MIPS $\lambda$24 {\micron} band, as shown in Figure 1, however, MGE028.4451$+$00.3094 is also weakly detected in the MIPS $\lambda$70 {\micron} band.  The nature of their nebular spectral energy distribution (SED) is uncertain. However, the lack of counterparts in any of the shorter wavelength IRAC bands suggests that there is no significant emission from warm dust ($T\gtrsim200$ K) or polycyclic aromatic hydrocarbons (PAH). The nebulae have $\lambda$24 {\micron} average surface brightnesses of $\approx$50 MJy sr$^{-1}$ along their rims and reach $\approx$180 MJy sr$^{-1}$ at their brightest sections. To be this bright at $\lambda$24 {\micron}, yet below the IRAC sensitivity limit, implies that any dust present must be cool ($T\lesssim100$--125 K). The weak detection of MGE028.4451$+$00.3094 at $\lambda$70 {\micron} is consistent with the presence of cool dust. This appears to be a common feature of the circumstellar nebulae that have been found around hot stars. The mid-infrared SED of the M1-67 nebula is dominated by a cool dust component having $T\approx125$ K, in addition to weaker, warm dust components with  $T\approx$200--500 K (Morris et al., in prep). The SEDs of the nebular material associated with the WNh stars WR102ca and WR102ka in the Galactic center also exhibit strong thermal continua from dust having $T\approx$130-200 K (\ca{barn08}~\cy{barn08}). Finally, the nebulae surrounding the LBVs WRA 751 and AG Car, and the P-Cygni star HDE316285, are also dominated by cool dust, with $T\approx75$--100 K (\ca{voors2000}~\cy{voors2000}; \ca{mor08}~\cy{mor08}). Analogous to these known examples, cool dust emission could provide a natural explanation for the MIPS $\lambda$24 {\micron} and $\lambda$70 {\micron} fluxes of MGE028.4812$+$00.3368 and MGE028.4451$+$00.3094.

It is also worth speculating on the contribution of line emission to the nebular SEDs.  Which atomic transitions will be prominent is dependent on the chemistry of both the current stellar wind and the mass lost in prior phases, and on the ionization temperature of the gas, which is a function of the radiation field that illuminates the nebula and the momentum of the stellar winds that impact it. Although morphologically distinct, the M1-67 nebula has a central WN8h star that is very similar to WMD47$^*$ and WMD48$^*$, so, the stellar wind and ionizing radiation field impinging on MGE028.4812$+$00.3368 and MGE028.4451$+$00.3094 should be comparable to the case of M1-67. \textit{Spitzer}/IRS spectra of M1-67 show that the MIPS $\lambda$24 {\micron} bandpass contains low-excitation transitions of [Fe {\sc ii}] and [Fe {\sc iii}] at $\lambda\lambda$22--26 {\micron}; yet, the total flux from these lines is relatively minor, compared with the thermal continuum flux (Morris et al., in prep).  We might conclude this is also the case for MGE028.4812$+$00.3368 and MGE028.4451$+$00.3094. However, several other ionizing stars exist in the vicinity of these nebulae, including a WC8 star that is presumably much hotter ($T_{eff}\approx80$ kK;  \ca{cro07}~\cy{cro07}) than the WN8--9h stars ($T_{eff}\approx30$--40 kK), with a faster and more chemically-enriched stellar wind. The \textit{Spitzer}/IRS spectra of the G24.1$+$1.4 nebula, which surrounds the hot WO star WR102, may provide an example of the effects of a very hot ionization source; it exhibits a nebular spectrum that is completely dominated by [O {\sc iv}] emission at $\lambda$25.89 {\micron}, in addition to [N {\sc v}] at $\lambda$24.3 {\micron} (Morris et al. in prep). This line is also prominent in the mid-infrared spectra of high-excitiation planetary nebulae (\ca{pmor06}~\cy{pmor06}; \ca{fes10}~\cy{fes10}; \ca{chu09}~\cy{chu09} and references therein).  It is interesting in this regard that the brightest halves of the MGE028.4812$+$00.3368 and MGE028.4451$+$00.3094 nebulae face in the approximate direction of the WC8 star, as shown in Figure 1, which might be an indication that line emission is contributing to the enhanced nebular brightness. Finally, although we suspect that the MIPS $\lambda$70 {\micron} emission from MGE028.4451$+$00.3094 is most likely the result of a cool dust continuum, we note that there are atomic transitions of [N {\sc iii}] at $\lambda$57.3 {\micron}, [O {\sc iii}] at $\lambda$88.4 {\micron}, and [O {\sc i}] at $\lambda$63 {\micron}, all of which have been observed in hot-star nebulae (\ca{moor80}~\cy{moor80}; \ca{voors2000}~\cy{voors2000}); their potential contribution cannot be completely ruled out.  

A particularly interesting feature of the MGE028.4812$+$00.3368 and MGE028.4451$+$00.3094 nebulae is their partial overlap on the sky. The outer rim of a circle centered on WMD47$^*$, illustrated in Figure 4, intersects a bright ridge of $\lambda$24 {\micron} emission having a curvature that is also centered on WMD47$^{*}$; we call this feature \textit{Ridge 1}. The feature is relatively bright in the MIPS $\lambda$70 {\micron} image as well, as seen in Figure 1\footnote{The MIPS $\lambda$70 {\micron} image was filtered to remove detector artifacts. Thus, although the image includes useful morphological information, reliable flux values were not preserved after the filtering process.}. The peak surface brightness of Ridge 1 at $\lambda$24 {\micron} is $\approx$140 MJy sr$^{-1}$, which is significantly brighter than the sum of the average rim surface brightnesses of MGE028.4812$+$00.3368 and MGE028.4451$+$00.3094 ($\lesssim$50 MJy sr$^{-1}$), away from the overlapping region. Thus, the brightness enhancement of Ridge 1 might not  be a simple projection effect,  but could actually be an intrinsically brighter portion of the nebula. It is worth speculating that Ridge 1 might actually be a region of contact between MGE028.4812$+$00.3368 and MGE028.4451$+$00.3094, a possibility noted in \ca{wac10}~(\cy{wac10}) and \ca{gvar10}~(\cy{gvar10}). Such contact, however, would have to be at an early stage, in order to have not disturbed the overall high degree of spherical symmetry for both nebulae.

The O7--8 III--I star 2MASSJ 18420957$-$0350313 also appears to be dramatically influencing the nebular morphology of MGE028.4451$+$00.3094, but on a smaller spatial scale. This star lies near the center of curvature of a bright ridge of $\lambda$24 {\micron} emission within MGE028.4451$+$00.3094. This feature, which we call \textit{Ridge 2}, has the highest surface brightness anywhere on the nebula ($\approx$180 MJy sr$^{-1}$). It is brightest on the side facing the WN8--9h star WMD48$^*$, which indicates that this feature is likely to be the result of an interaction between the wind of the O7--8 III--I star and the outflow from WMD48$^*$. 

Finally, the VLA $\lambda$20 cm MAGPIS survey image (\ca{hel06}~\cy{hel06}) in Figure 4 (\textit{left panel}) exhibits a faint shell that appears to trace the inner edge of MGE028.4451$+$00.3094 along its eastern side. This could be where the supersonic winds of the WN8-9h star WMD48$^*$ and the O7--8 III--I star have shocked and ionized the inner edge of the shell. As such, the $\lambda$20 cm emission could be free-free continuum or non-thermal synchrotron.  The radio flux on the eastern side of the nebulae suggests the presence of H {\sc i} from the neighboring ISM, which also coincides with the diffuse $\lambda$8.0 {\micron} emission (probably PAH), apparent in this part of the sky in Figure 1. The lack of diffuse $\lambda$20 cm or  $\lambda$8.0 {\micron} emission on the opposite (western) sides of the nebulae suggests that MGE028.4812$+$00.3368 and MGE028.4451$+$00.3094 might be pushing into denser ISM on their eastern sides. As such, the arc of enhanced radio emission on the eastern side of MGE028.4451$+$00.3094 could be the result of compressed ISM.

The MGE028.4812$+$00.3368 and MGE028.4451$+$00.3094 system and its association of evolved massive stars represents an unprecedented display of interactions between multiple stellar outflows, radiation fields, and the ISM. The situation is undoubtedly too complex to be accurately described by our qualitative assessment of the images and our comparison with known hot-star nebulae.  Mid-infrared spectroscopy with the \textit{Herschel Space Telescope} and subsequent spectral modeling could elucidate the nature of the nebular emission, and untangle the flux contributions of thermal dust and emission lines. This would allow for an estimate of the nebula masses.  Finally, although no WR candidates remain in the near vicinity, at least not according to the color selection criterion of \ca{had07}~(\cy{had07}) and \ca{mau09}~(\cy{mau09}), it is very likely that there are additional massive stars that remain unidentified in this region, which might also be affecting the dynamics and emission processes responsible for MGE028.4812$+$00.3368 and MGE028.4451$+$00.3094. 

\begin{acknowledgements}
We thank the referee Dr. Wolf-Rainer Hamann for helping to improve this manuscript. This research is based on observations made at the Hale Telescope, Palomar Observatory, a continuing collaboration between Caltech, NASA/JPL, and Cornell University.
\end{acknowledgements}

\begin{deluxetable}{lrrrr}
\tablecolumns{5}
\tablewidth{0pc}
\tabletypesize{\scriptsize}
\renewcommand{\arraystretch}{1.4}
\tablecaption{Photometry of the Massive Stars}
\tablehead{ \colhead{Star} & \colhead{WMD47$^*$, 2MASSJ } & \colhead{WMD48$^*$, 2MASSJ } & \colhead{2MASSJ } & \colhead{2MASSJ } \\ [2pt]
\colhead{} &  \colhead{18420630$-$0348224} & \colhead{18420827$-$0351029}  & \colhead{18420846$-$0349352}  & \colhead{18420957$-$0350313}}
\startdata
Spectral Type               & WN8--9h  & WN8--9h & WC8  &  O7--8 III--I \\
R.A. (J2000)                   &  280.526285 & 280.534464 & 280.535279 & 280.539911 \\
Decl. (J2000)                & $-$3.806245 & $-$3.850808 &$-$3.826450 & $-$3.842041\\
$J$ (mag)      & 11.95 (0.03) & 11.85 (0.03) & 11.93 (0.02) &12.39 (0.03)  \\
$H$ (mag)     &10.22 (0.02) & 10.26 (0.03)& 10.83 (0.02)&10.61 (0.03) \\
$K_s$ (mag)& 9.16 (0.02) & 9.27 (0.03) & 9.75 (0.02) & 10.01 (0.03) \\
${(J-K_s)}_0$ & 0.13 & 0.13 & 0.43 & $-$0.21 \\
${(H-K_s)}_0$ & 0.11 & 0.11 & 0.38  & $-$0.11 \\
$A^{J-K_{s}}_{K_{s}}$ & 1.78 (0.19) & 1.64 (0.17) & 1.17 (0.12) & 1.74 (0.18) \\ 
$A^{H-K_{s}}_{K_{s}}$& 1.72 (0.29) & 1.61 (0.27) & 1.28 (0.21) & 1.30 (0.23) \\ 
$\overline{A_{K_{s}}}$&1.75 (0.34) & 1.62 (0.32) & 1.23 (0.25) & 1.52 (0.29)  \\ 
$M_{K_s}$ (adopted)& $-$5.9 & $-$5.9 & $-$4.9 &  $-$5.0\tablenotemark{a}\\
D (kpc)& 4.6 (0.7) & 5.2 (0.8) & 4.8 (0.6) & 4.9 (1.2) 
\enddata
\tablecomments{Uncertainties are within parentheses. All 2MASS photometric measurements were obtained on Julian date 2451312.8338. Intrinsic colors and $M_K$ values were adopted from \ca{cro06}~(\cy{cro06}) for Wolf-Rayet stars, and from \ca{mar06}~(\cy{mar06}) for the O star (colors only, no $M_K$).}
\tablenotetext{a}{Since $M_K$ could not be adopted for this source (O star of uncertain luminosity class), we derived $M_K$ by using the extinction for this source and the average distance for the three Wolf-Rayet stars in the table ($4.9\pm1.2$ kpc).}
\end{deluxetable}
\newpage

\begin{figure*}[t]
\centering
\epsscale{1}
\plotone{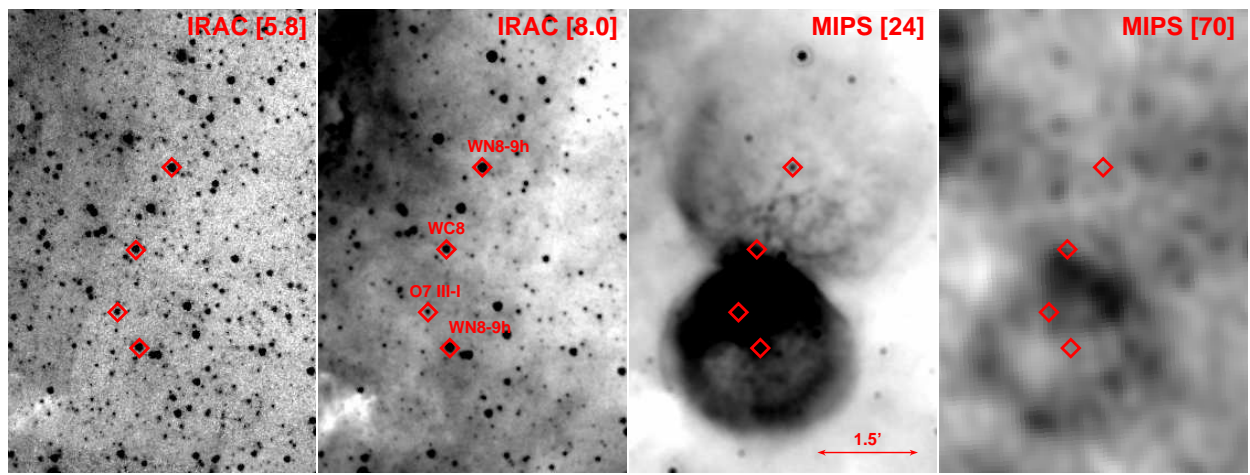}
\caption[]{\linespread{1}\normalsize{Mid-infrared field images containing the MGE028.4812$+$00.3368 and MGE028.4451$+$00.3094 nebulae.  All images are aligned to the same orientation and scale. The associated massive stars are marked with diamond symbols in each image, and their spectral types are labeled in the IRAC [8.0] image. North is up and east is toward the left.}}
\end{figure*}

\begin{figure*}[t]
\centering
\epsscale{1}
\plotone{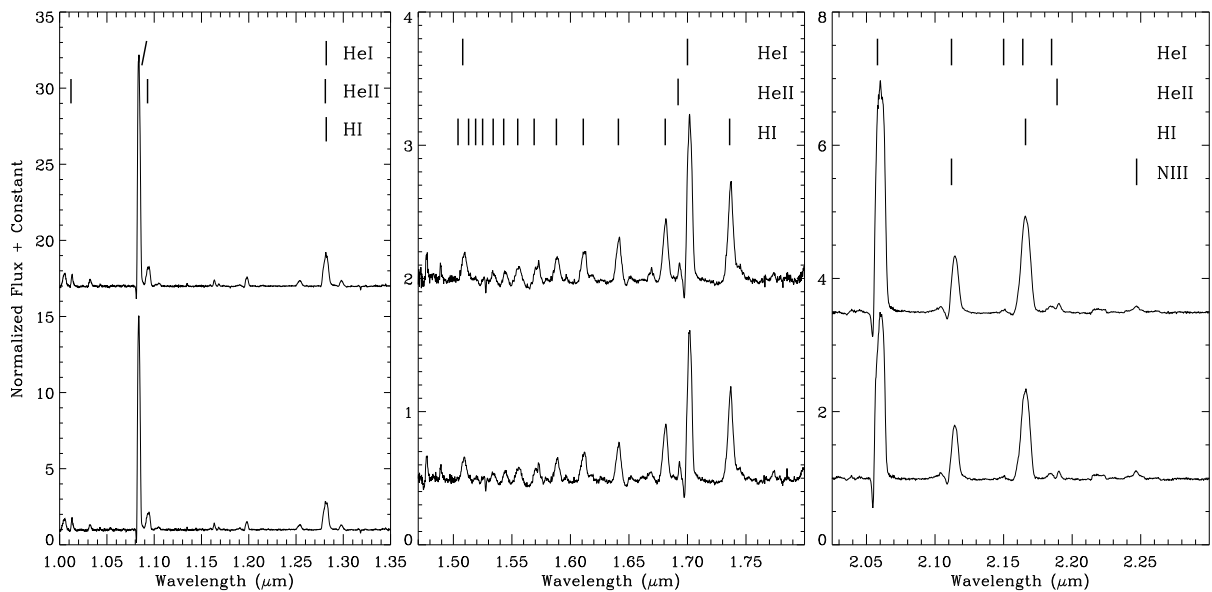}
\caption[]{\linespread{1}\normalsize{$JHK$ spectra of WMD47$^*$ (\textit{upper spectrum}) and WMD48$^*$ (\textit{lower spectrum}), obtained with the Palomar Hale 5 m telescope and Triplespec spectrograph. The spectral morphologies of both stars are remarkably similar to each other and indicate WN8--9h spectral type.}}
\end{figure*}

\begin{figure*}[t]
\centering
\epsscale{1}
\plotone{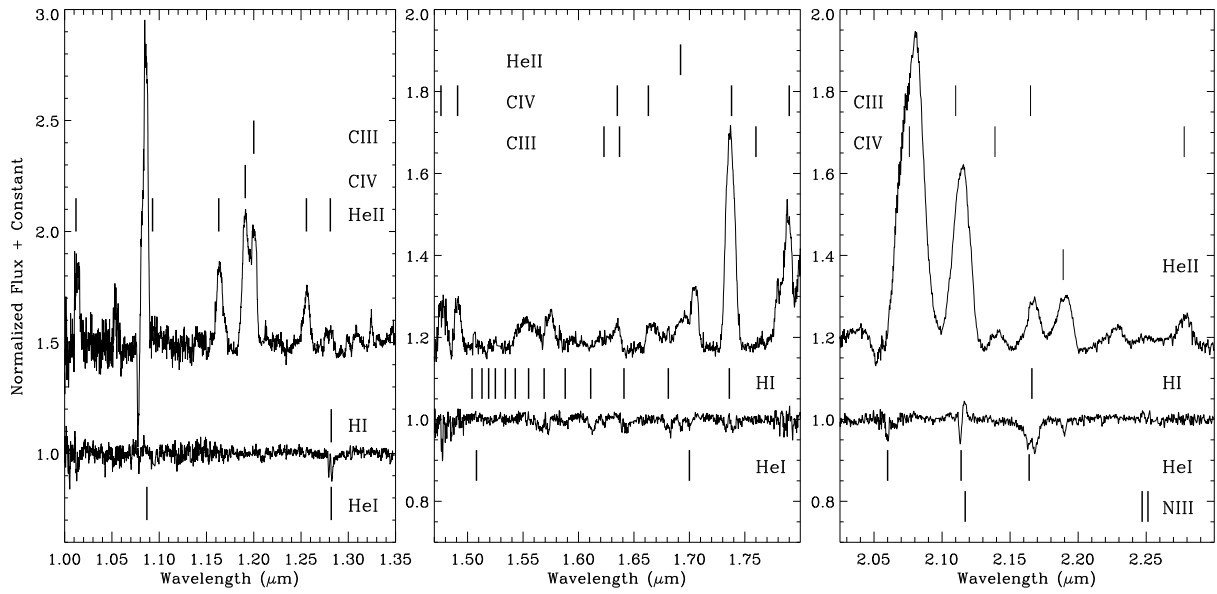}
\caption[]{\linespread{1}\normalsize{$JHK$ spectra of the WC8 star 2MASSJ 18420846$-$0349352 (\textit{upper spectrum}) and O7--8 III--I star 2MASSJ 18420957$-$0350313 (\textit{lower spectrum}).}}
\end{figure*}

\begin{figure*}[t]
\centering
\epsscale{1}
\plotone{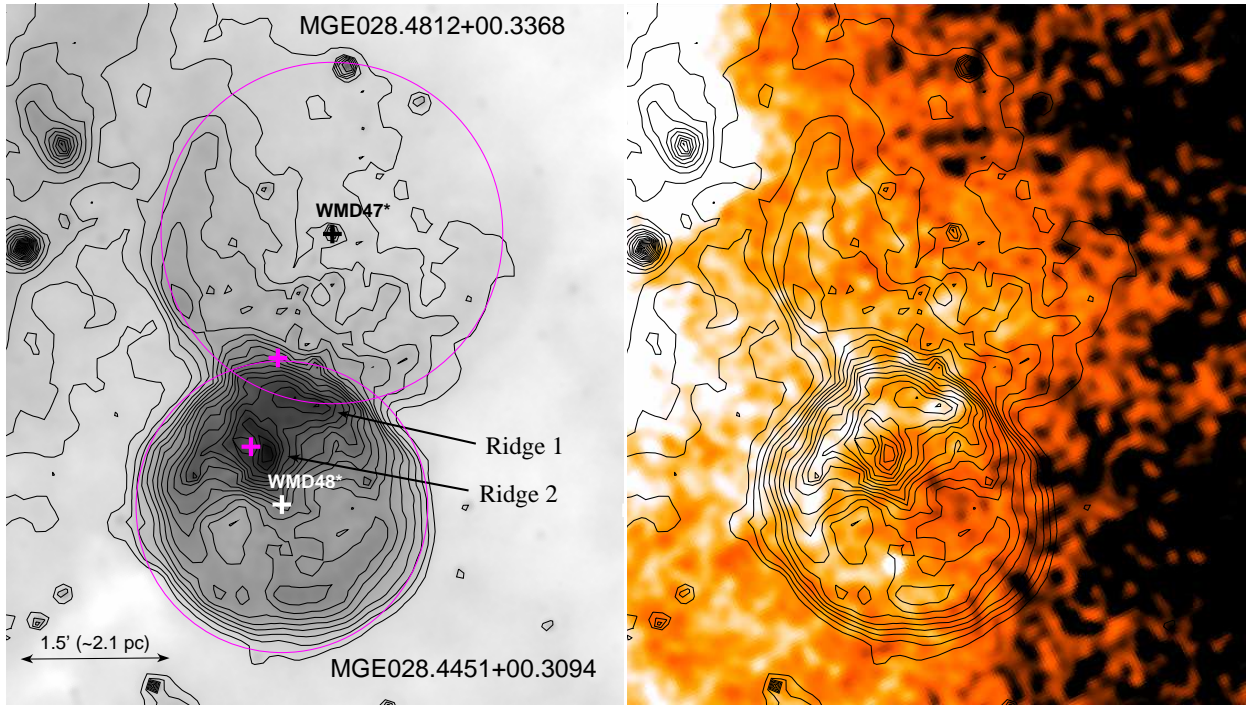}
\caption[]{\linespread{1}\normalsize{\textit{Left}: \textit{Spitzer}/MIPS $\lambda$24 {\micron} image of MGE028.4812$+$00.3368 and MGE028.4451$+$00.3094, with logarithmic brightness contours. The positions of the WN8--9h stars WMD47$^*$ and WMD48$^*$ (\textit{white and black crosses}) and the WC8 (\textit{upper magenta cross}) and O7--8 III--I (\textit{lower magenta cross}) stars are marked. Ridge 1 is concentric with the circle that is centered on WMD47$^*$, and is delineates the region where the two nebulae intersect. Ridge 2 envelops the O7--8 III--I star (see text) and was likely created by its wind. \textit{Right}: False color VLA $\lambda$20 cm radio image from MAGPIS survey (\ca{hel06}~\cy{hel06}) with the same \textit{Spitzer}/MIPS $\lambda$24 {\micron} contours as before. Note the ring of radio emission that traces the inner edge of the MGE028.4451$+$00.3094 nebula, particularly on its eastern side. North is up and east is toward the left.}}
\end{figure*}

\end{document}